\documentclass[fleqn,10pt]{wlscirep}
\usepackage{multirow} 
\usepackage{siunitx}
\usepackage{makecell}
\usepackage{nicematrix}
\usepackage{array}
\usepackage[utf8]{inputenc}
\usepackage[T1]{fontenc}
\usepackage{pdfpages}
\usepackage{lineno}
\usepackage{tabulary}
\usepackage{upgreek}
\usepackage{hyperref}

\title{Measured group birefringence and group velocity dispersion of elliptical-core ZBLAN fibres for mid-infrared supercontinuum generation}

\author[1,2,*]{Shreesha Rao D. S.}
\author[1,3]{Christian R. Petersen}
\author[2]{Anupamaa Rampur}
\author[1,3]{Ole Bang}
\author[2]{Alexander M. Heidt}

\affil[1]{DTU Electro, Department of Electrical and Photonics Engineering, Technical University of Denmark, {\O}rsteds Plads, 2800 Kongens Lyngby, Denmark}
\affil[2]{Institute of Applied Physics, University of Bern, Sidlerstrasse 5, 3012 Bern, Switzerland.}
\affil[3]{NORBLIS ApS, {\O}rsteds Plads 343, 2800 Kongens Lyngby, Denmark.}
\affil[*]{\small corresponding author: Shreesha Rao D. S. (shdes@dtu.dk)}
\begin{abstract}
Polarisation-maintaining ZBLAN optical fibres with small elliptical cores are attractive for nonlinear frequency conversion, for example in supercontinuum generation (SCG), and for dispersion management in ultrafast fibre laser systems operating in the mid-infrared (MIR) spectral region. Accurate characterisation of group birefringence and group velocity dispersion (GVD) is essential for modelling ultrafast pulse propagation in such fibres, yet experimental data remain scarce. Here, we present broadband measurements of group birefringence and GVD in four elliptical-core ZBLAN fibres, selected for their potential for low-noise MIR SCG, over the wavelength range 1.4 to 4.3~$\upmu$m. Using polarisation-resolved white-light spectral interferometry, we quantify the dispersion characteristics of both fundamental modes in each fibre. Measurement uncertainties are evaluated using a combined uncertainty analysis that includes the spread between repeated measurements and the uncertainty in the measured fibre length. This work aims to support the development of MIR supercontinuum sources and to facilitate the understanding of nonlinear optical phenomena in ZBLAN fibres.
\end{abstract}
\begin{document}
\flushbottom
\maketitle
\section*{Background \& summary}
Optical fibres are critical components in modern science and technology, serving applications such as telecommunications, environmental sensing, and laser systems. Their performance depends on the ability to accurately control both linear and nonlinear light propagation, which in turn depends on the optical properties of the fibre. Accurate modelling of pulse propagation is essential to predict how light interacts with the waveguide structure and material dispersion. Such modelling not only provides insight into the underlying physics but also enables accurate design and fabrication of fibres to achieve desired characteristics for specific applications.

Optical fibres are waveguides made of a core and cladding, designed to confine and guide light. The refractive index of the core material is denoted n$_{core}$, and that of the cladding as n$_{clad}$. Fibres can be engineered to support only the fundamental mode for frequencies lower than a certain cut-off frequency. In addition, some fibres can be designed to be endlessly single mode. The modes that propagate in the fibre experience an effective refractive index n$_{eff}(\upomega)$, which depends on frequency and accounts for the waveguide structure and material properties. The propagation constant, $\upbeta(\upomega)$, is related to n$_{eff}(\upomega)$ as $\upbeta(\upomega) =$ n$_{eff}(\upomega)\cdot\upomega\mathbin{/}$c, where $\upomega$ is the angular frequency of light, and c is the speed of light in vacuum. The first derivative of the propagation constant, $\upbeta_{1}(\upomega) =$ d$\upbeta/d\upomega$, gives the group velocity v$_{g}$ of an optical pulse, with v$_{g}=$1$\mathbin{/}\upbeta_{1}(\upomega)$. The second derivative, $\upbeta_{2}(\upomega)=d^2\upbeta\mathbin{/}d\upomega^2$, is the  group velocity dispersion (GVD), a critical parameter that determines the extent to which different frequency components of a pulse spread temporally as they propagate. Group velocity dispersion, $\upbeta_{2}(\upomega)$, is typically expressed in units of ps$^{2}\mathbin{/}$km. Another commonly used group velocity dispersion parameter is D, which is defined as the derivative of $\upbeta_{1}(\upomega)$ with respect to wavelength, $\uplambda$, and is usually expressed in units of ps$\mathbin{/}$(km$\cdot$nm). The relation between the two group velocity dispersion parameters is given by D $= -$2 $\uppi$ c $\cdot \upbeta_{2}\mathbin{/}\uplambda^{2}$.

The GVD determines how the pulse evolves during propagation. It is a key linear effect that influences pulse propagation in the fibre. As the pulse propagates through the fibre, its constituent frequency components can become temporally separated due to differences in their propagation speeds. This frequency variation across the pulse duration is referred to as chirp. If the intensity of the pulse travelling through the fibre is high enough, nonlinear effects may arise. The pulse then experiences an interplay between linear, and nonlinear effects. One such nonlinear effect is self-phase modulation (SPM), which occurs when the refractive index experienced by the pulse depends on its intensity. As the intensity of the pulse varies along its duration, the pulse experiences a nonlinear refractive index change over the pulse duration, leading to the generation of new frequencies. As a result, SPM can induce a chirp across the pulse due to this nonlinearity. In the special case where the nonlinear chirp from SPM, and the linear chirp from GVD counteract each other, the pulse can evolve into a stable state which maintains its spectral and temporal profile during further propagation. Such pulses, which preserve their spectro-temporal characteristics over long distances due to a precise balance between dispersive and nonlinear effects, are known as solitons. The pulse propagation in the fibre can be modelled using the generalised nonlinear Schr\"{o}dinger equation (GNLSE)~\cite{Agr19NoFi, Dudl10SupGen}. This equation accounts for both the linear and nonlinear effects influencing pulse evolution in the fibre. A key parameter in implementing the GNLSE is the wavelength-dependent GVD profile of the fibre, which must be precisely known for accurate modelling of pulse evolution.

GVD measurement across a broad wavelength region is particularly important in studies of supercontinuum generation (SCG) in fibres. SCG is an extreme nonlinear process which leads to a spectrally broad output in a medium, when pumped by a relatively narrow-band input pulse. In a traditional SC source, a nanosecond (ns) or a picosecond (ps) pump with a central wavelength slightly longer than the fibre's zero-dispersion wavelength (ZDW) is launched into the fibre. The spectrum primarily broadens through modulation instability, which leads to soliton formation in the anomalous dispersion regime and the generation of associated dispersive waves in the normal dispersion regime~\cite{Isl89OlBrB}. These dispersive waves contribute significantly to the spectral extension of the SC. The wavelength range at which dispersive waves are generated can be accurately modelled when the fibre's GVD is known across the entire spectral regime covered by the SC. With high-power ns and ps lasers around 1~$\upmu$m, and silica fibres with a ZDW at shorter wavelengths than the pump, SC spanning 390--2400~nm can be achieved~\cite{Wad04OeBSc}. This range covers the full transmission window of silica fibres. The ability to model and optimise such broad spectra depends critically on accurate measurements of GVD across the entire bandwidth. These multi-octave-spanning sources have been used in confocal fluorescence microscopy~\cite{McC07Tim, Jal08Sup}, among other applications.

A major drawback of such SC sources is their high pulse-to-pulse fluctuations~\cite{Nak98OftCDeg}. This is inherent to the fundamental mechanism driving SCG. Since modulation instability is seeded by noise, the temporal and spectral profiles can vary significantly from pulse-to-pulse, even when the pump pulse characteristics remain unchanged~\cite{RevM06ScDud}. One way to overcome this limitation is to generate the SC entirely in the normal dispersion regime of the fibre~\cite{Het10FltTp}. When a femtosecond (fs) pulse is launched into the fibre in the normal dispersion regime, the coherent processes of SPM and optical wave breaking (OWB) can produce an ultra-low-noise SC~\cite{Klim16SrDft}, provided the entire generated spectrum remains within the normal dispersion regime~\cite{Ull19ShrOl}. Such an SC can maintain a single pulse in the time domain and a flat, smooth spectral profile~\cite{Het10FltTp}. 
 
A single-mode fibre can support two fundamental modes that correspond to orthogonal polarisation states. In a low-birefringence fibre, these two orthogonal modes have very similar characteristics; for example, their group velocities can be nearly identical. As the low-birefringence results in similar propagation constants for the two modes, light can couple between them during propagation. As the group velocities of these modes are similar over a broad wavelength range, power in one mode can noise-seed the spectrum in the other, leading to polarisation mode instability (PMI) -- a nonlinear coupling phenomenon causing power transfer between polarisation modes. Since the spectra resulting from PMI are noise-seeded, the resulting SC can exhibit high pulse-to-pulse fluctuation~\cite{SrPmi18Ivn}, despite being in the normal dispersion regime. By incorporating stress rods in the cladding, shaping the core elliptically, combining both methods, or by using other methods, for example, by structuring the cladding near the core to break symmetry, the phase birefringence of the fibre can be increased. This ensures that the propagation constants of the two polarisation modes are sufficiently separated; these fibres are known as polarisation-maintaining (PM) fibres. While phase birefringence determines the separation in propagation constants between the two modes, PM fibres require sufficiently high group birefringence to effectively suppress PMI. In fibres with high group birefringence, the large difference in group velocities prevents coupling between the polarisation modes~\cite{SaVMI05Bra}. Using such PM fibres with a normal dispersion profile enables the generation of ultra-low-noise SC~\cite{SrPmi18Ivn, Rao22PmNpm}, which is essential for applications requiring high sensitivity in detection.

Ultra-low-noise SC sources, exhibiting relative-intensity noise (RIN) levels comparable to those of technologically mature ultra-low-noise fs pump sources at 1.0~$\upmu$m~\cite{Eti21ExpOl} and 1.55~$\upmu$m~\cite{Ben22OptNoi}, have been demonstrated in silica-based fibres. However, their spectral extension has typically been limited to the near-infrared (NIR), up to approximately 2.4~$\upmu$m, due to the high loss in silica fibres at longer wavelengths. These SC sources enable a number of applications in the NIR such as high-quality pulse compression~\cite{Hei11OpEPuC}, single-beam coherent anti-Stokes Raman spectroscopy~\cite{Liu13OpECars}, ultra-high resolution (UHR) spectral-domain optical coherence tomography (OCT)~\cite{Shr21Shot}, UHR apertureless scanning near-field optical microscopy~\cite{Kob21Near} (SNOM), and dual-comb spectroscopy~\cite{Gru24DuaOl}. 

Extending SCG into the mid-infrared (MIR) region is highly desirable for a number of applications, for example in spectral-domain OCT~\cite{Nie19LsaMirO}. In spectral-domain OCT, penetration depth is limited when using NIR sources due to scattering losses, which for small particles scale approximately as $\uplambda^{-4}$. As scattering decreases rapidly with increasing wavelength in this regime, using a MIR source can significantly increase penetration depth. Fluoride fibres are excellent candidates to extend SCG into the MIR due to their broad transmission window, which spans from below 300~nm to beyond 4.5~$\upmu$m, thereby covering the visible, NIR, and a part of MIR regions. Among these, one of the fibres that are commercially available is made from ZBLAN (ZrF$_{4}$--BaF$_{2}$--LaF$_{3}$--AlF$_{3}$--NaF). To date, SCG in ZBLAN fibres has been demonstrated by pumping in the anomalous dispersion regime~\cite{Hag06ZbScIe,Xia06ZbScOl}, and its applications are beginning to be explored using such sources~\cite{Nie19LsaMirO}. An SC spanning this MIR region also covers important gas absorption features, including carbon monoxide~\cite{PlyCoE58Jrnbs}, methane~\cite{PlyCh458Jrnbs}, ammonia~\cite{BenNH358Jrnbs}, and other molecules, making it highly relevant for spectroscopy and sensing. However, measured dispersion data for ZBLAN fibres remain scarce, limiting accurate modelling and optimisation of SC sources. Furthermore, SCG in normal-dispersion fluoride fibres has not yet been studied, largely due to the lack of fibres with such dispersion profiles. Ultra-low-noise SCG in PM-ZBLAN fibres with tailored normal dispersion is needed for a range of applications; for instance, it could significantly improve the sensitivity of MIR OCT systems~\cite{Nie19LsaMirO,Shr21Shot} and enable SNOM in the MIR, which is yet to be realised because suitable ultra-low-noise sources are not available.

By providing broadband measurements of both group birefringence and GVD for the two fundamental modes of each fibre, this work fills an important gap in the literature. The wavelength range covered, 1.4 to 4.3~$\upmu$m, encompasses the majority of the low-loss transmission region of ZBLAN and extends well beyond the range normally accessible using conventional white-light interferometry, which is typically limited to the transmission window of silica fibres.

\begin{figure}[htbp!]
\centering
\fbox{\includegraphics[width=0.94\linewidth]{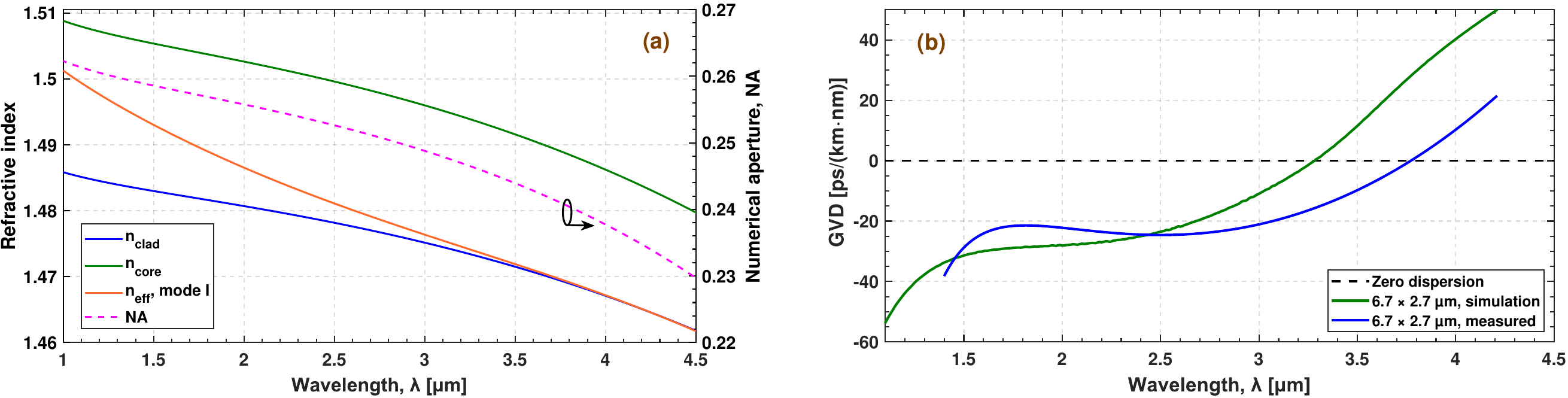}}
\caption{Representative simulation of the 6.7~$\times$~2.7~$\upmu$m core elliptical PM-ZBLAN fibre. (a) Wavelength-dependent core and cladding refractive indices from the manufacturer of the ZBLAN fibres with nominal NA $=$ 0.26 at 1.24~$\upmu$m, the corresponding NA, and the simulated effective refractive index n$_\mathrm{eff}$ of one of the fundamental modes. (b) Simulated dispersion obtained from the finite-element-method calculation, and the measured dispersion of mode~I of the 6.7~$\times$~2.7~$\upmu$m core fibre.}
\label{Fig:SimDisp}
\textcolor{color1}{\dotfill}
\end{figure}
An important motivation for these measurements is that simulations based on nominal fibre parameters do not necessarily reproduce the measured dispersion characteristics of ZBLAN fibres. To illustrate this limitation of relying on nominal fibre parameters, we performed a representative finite-element-method calculation for the 6.7~$\times$~2.7~$\upmu$m core fibre investigated in this study. The calculation used the wavelength-dependent core and cladding refractive indices provided by the manufacturer for a ZBLAN fibre with a nominal numerical aperture (NA) of 0.26 at 1.24~$\upmu$m, corresponding to the nominal NA of the 6.7~$\times$~2.7~$\upmu$m core fibre. The core and cladding refractive indices, the corresponding NA, and the calculated effective refractive index of one of the fundamental modes are shown in Fig.~\ref{Fig:SimDisp}(a). The simulated dispersion calculated from this mode and the measured dispersion of mode~I are plotted in Fig.~\ref{Fig:SimDisp}(b).  Although the simulation gives a useful qualitative reference, it does not agree quantitatively with the measured GVD over the full wavelength range. A rigorous quantitative simulation would require accurate knowledge of the wavelength-dependent core and cladding refractive indices of the actual glasses used in each fibre across the transmission range. Such information is not available with sufficient certainty for the off-the-shelf commercial samples studied here. Moreover, the four fibres have different NAs, indicating different index contrasts, and the dispersion may also be influenced by fabrication-dependent effects such as residual stress. In fluoride glass fibres, changes in refractive index during fabrication have been attributed to changes in glass structure during drawing and to stress induced during preform fabrication~\cite{NakCRi90Apl}. These effects can locally modify the refractive index, particularly near the core--cladding interface, and may depend strongly on the fabrication method. Consequently, deviations between nominally simulated and experimentally realised fibre properties are expected. This limited reliability of purely predictive modelling is one of the main motivations for the direct broadband measurements reported here. The measured GVD values also provide an experimental reference point for earlier simulation-based studies of ZBLAN fibres, including studies on circular-core ZBLAN fibres~\cite{Kub13ZbDJosaB}.

The measured dispersion values correspond to the actual off-the-shelf fibres studied here, making this dataset a practical starting point for initial simulations and experimental planning. This is particularly useful when detailed fibre-specific refractive-index and fabrication data are unavailable, and when dispersion calculated from nominal parameters does not accurately reproduce the experimentally realised fibre properties. Therefore, for researchers seeking to model or optimise MIR nonlinear propagation in these fibres, the experimentally measured dispersion curves reported here provide a more reliable starting point than dispersion obtained only from nominal fibre simulations. The results therefore demonstrate what is practically achievable in terms of GVD in commercially available elliptical-core PM-ZBLAN fibres. The elevated group birefringence achievable in these fibres suppresses PMI over a broad spectral range, thereby enabling the generation of ultra-low-noise SC entirely within the normal dispersion regime. The range of core dimensions studied encompasses PM-ZBLAN fibres with ZDW shorter than 2~$\upmu$m to longer than 3.75~$\upmu$m. In addition to their relevance for MIR ultra-low-noise SC sources based on PM-ZBLAN fibres---a capability that has not yet been fully demonstrated---accurate knowledge of the dispersion and group birefringence of these fibres is valuable for a broad range of applications such as third-harmonic generation~\cite{Gao13ElZbOl} and dispersion management in 2.8~$\upmu$m laser systems~\cite{Yan90Zb3El}. The measured group birefringence and GVD data provided in this paper are intended to support experimental planning, reproducible modelling, reduce uncertainty in numerical studies, and facilitate the development of next-generation MIR fibre sources.

\section*{Methods}
The fibres were cleaved using a Vytron cleaver (Thorlabs: CAC4020AFJ) with a tension of $\sim$100~g. The intended fibre length after cleaving was approximately 20 cm. After cleaving the input and output ends, the actual length of each fibre was measured using a ruler with an accuracy of $\pm$1~mm. The fibres were mounted with the input end held in a keyway bare fibre holder placed on a three-axis translation stage, and the output end secured in a keyway bare fibre rotation mount on a separate three-axis translation stage.  The sequence of measurements was as follows: (i) identification of the principal axes of the fibre, (ii) broadband measurement of the group birefringence, and (iii) broadband measurement of the GVD.

\begin{table}[htbp!]
\centering
\setlength{\extrarowheight}{0pt}
\setlength{\tabcolsep}{15pt}
\begin{NiceTabular}{c c c}[cell-space-limits=0pt]
\CodeBefore
\rowcolor{color3}{1}
\Body
\toprule
\textbf{Item name} & \textbf{Fibre core$\mathbin{/}$cladding [$\upmu$m]} & \textbf{Numerical aperture} \\
\midrule
ZEF-2.2$\times$5.5$\mathbin{/}$125-N & 5.5 $\times$ 2.4$\mathbin{/}$123 & 0.23 $\pm$ 0.02 at 1.20 $\upmu$m\\
\midrule
ZEF-2.7$\times$6.7 $\mathbin{/}$95$\times$125-N & 6.7 $\times$ 2.7$\mathbin{/}$124 $\times$ 96 & 0.26 $\pm$ 0.02 at 1.24 $\upmu$m\\
\midrule
ZEF-4.5$\times$9$\mathbin{/}$125-N & 8.9 $\times$ 4.1$\mathbin{/}$121 & 0.20 $\pm$ 0.02 at 0.84 $\upmu$m\\
\midrule
ZEF-6$\times$10$\mathbin{/}$125-N & 10.6 $\times$ 5.6$\mathbin{/}$121 & 0.20 $\pm$ 0.02 at 0.84 $\upmu$m\\
\bottomrule
\end{NiceTabular}
\caption{Core and cladding dimensions, along with NA values at specified wavelengths, for the elliptical core PM-ZBLAN fibres used in this study. These details are taken from the datasheets provided by the fibre manufacturer, FiberLabs Inc., Japan.} \label{Tab:Tab1}
\textcolor{color1}{\dotfill}
\end{table}

\begin{figure}[htbp!]
\centering
\fbox{\includegraphics[width=0.94\linewidth]{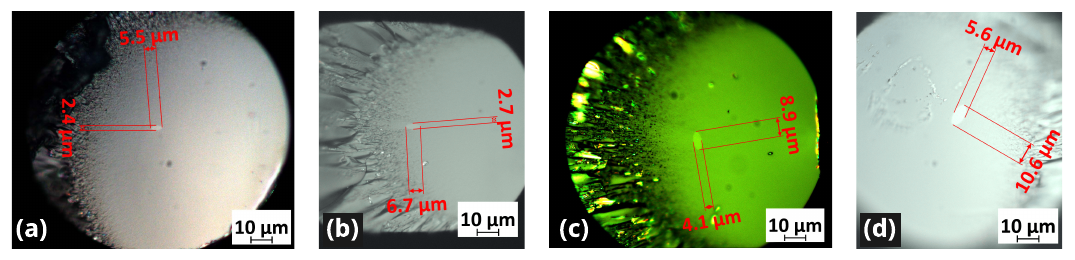}}
\caption{100$\times$ magnified cross-sections of the four elliptical-core fibres.}
\label{Fig:FibImg}
\textcolor{color1}{\dotfill}
\end{figure}
Four fibres were selected to represent a realistic and practically relevant subset of commercially available elliptical-core PM-ZBLAN fibres. The cross-sections of the four fibres are shown in Fig.~\ref{Fig:FibImg}(a)-(d). The core dimensions span from 5.5~$\times$~2.4~$\upmu$m to 10.6~$\times$~5.6~$\upmu$m. The core and cladding dimensions of the fibres are provided in Table~\ref{Tab:Tab1}, along with the NA values at specified wavelengths. These details are taken from the datasheets provided by the fibre manufacturer, FiberLabs Inc., Japan. The core dimensions are used to refer to the individual fibres. 

\subsection*{Measurement of group birefringence } 
\begin{figure}[htbp!]
\centering
\fbox{\includegraphics[width=0.98\linewidth]{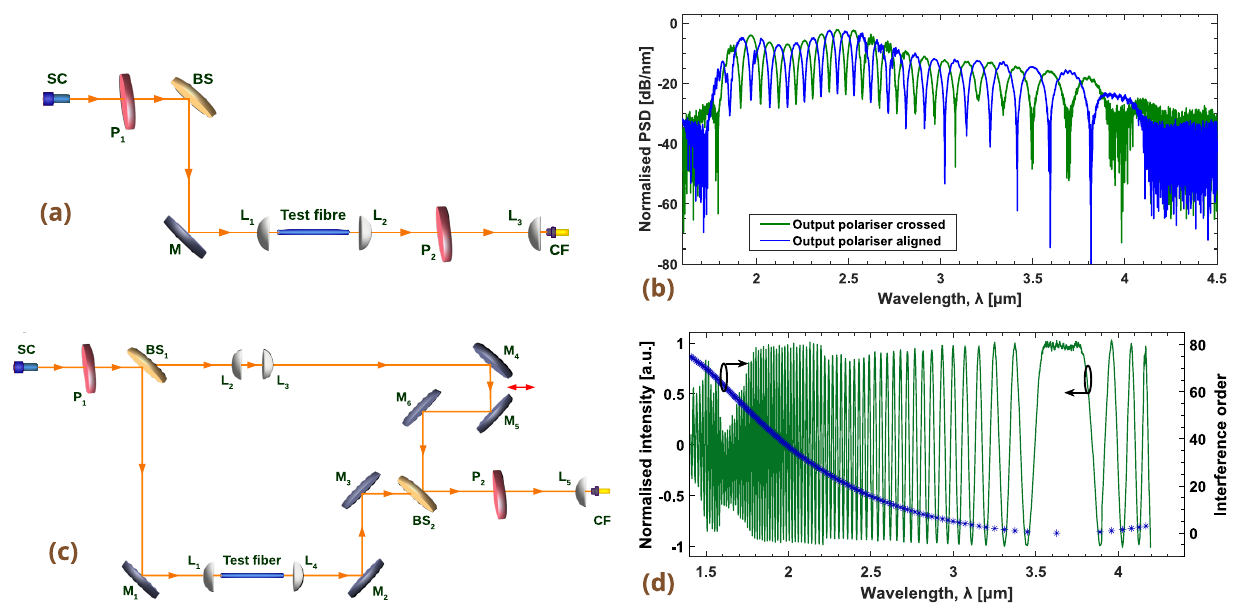}}
\caption{(a) Schematic of the group birefringence measurement setup. (b) Example interference spectra for the group birefringence measurement, with the output polariser crossed in green and aligned in blue relative to the input polariser. (c) Schematic of the white light spectral interferometry setup used for GVD measurement. (d) Example spectral interference pattern I$_{Interf}$, for the GVD measurement showing fringes around the phase-matching wavelength in green. The corresponding fringe order for this measurement is shown in blue with respect to the right-hand y-axis.}
\label{Fig:MetBAnD}
\textcolor{color1}{\dotfill}
\end{figure}

A source (NKT Photonics: SuperK MiR 15) covering the wavelength range 1.4--4.3~$\upmu$m was used. The source spectrum exhibits a peak at 1.55~$\upmu$m corresponding to the SC pump, and a dip around 1.7~$\upmu$m. The group birefringence measurement setup is shown in Fig.~\ref{Fig:MetBAnD}(a). For this measurement, only the reflected beam from the beam splitter (BS, Thorlabs: BSW511) was used. Light was coupled into the fibre using an aspherical lens L$_{1}$, and the output light was collimated with an aspherical lens L$_{2}$.  

The angular separation between the two fundamental modes of the fibre is typically 90$^\circ$, although their orientation may vary slightly between fibres due to structural asymmetries. To identify the angle of the transmission axis of the input polariser at which light is coupled into a given principal axis of the fibre, the transmission axis of the input polariser (P$_{1}$, Thorlabs: LPMIR050-MP2) was set to an arbitrary initial angle, say 0$^\circ$. Hereafter, `rotating the polariser' refers to rotating its transmission axis. At the start of the measurement, the output polariser (P$_{2}$, Thorlabs: LPMIR050-MP2) was rotated to locate the minimum transmitted power, and the corresponding angle, $\uptheta_{2,\min}$, was recorded; P$_{2}$ was then rotated to locate the maximum, and the angle $\uptheta_{2,\max}$ was recorded. At these angles, the transmitted power was measured.

The measurement proceeded by rotating P$_{1}$ in 10$^\circ$ steps; at each angle, the transmitted power through P$_{2}$ was recorded twice: first with P$_{2}$ set to $\uptheta_{2,\min}$, then with it set to $\uptheta_{2,\max}$. As P$_{1}$ was rotated through 360$^\circ$, each of the two traces varied near-sinusoidally with two maxima and two minima. The traces were complementary, with maxima and minima interchanged. The angles of P$_{1}$ at which the extrema occurred indicated the orientations that coupled light into one fundamental mode and into the orthogonal mode, respectively, based on polariser--analyser fibre characterisation methods~\cite{Pmd94PolJlt,Fol04LmaOe}. While either trace alone would suffice to find the extrema, both complementary traces were recorded at every angle to reduce any ambiguity when identifying the angles associated with coupling to the fibre's fundamental modes.

After the orientations of the two principal axes were determined, the input polariser P$_{1}$ was rotated such that equal power was coupled into both orthogonal fundamental modes of the fibre. When the angular separation between the maxima and the adjacent minima is 90$^\circ$, equal power can be coupled into the two fundamental modes by aligning P$_{1}$ at an angle of 45$^\circ$ to the principal axes. The light then passed through P$_{2}$ and was coupled to a collection fibre (CF, Thorlabs: P1-23Z-FC-1) using an aspherical lens (L$_{3}$, Thorlabs: C036TME-D), as shown in Fig.~\ref{Fig:MetBAnD}(a). The collection fibre was then connected to a broadband optical spectrum analyser (Instrument Systems: Spectro 320), capable of recording spectra in the wavelength range 0.18--5.0~$\upmu$m.

The spectra were recorded with the output polariser crossed~\cite{Fre1822,Ras82OpL} and then aligned with respect to the input polariser.  As the input polariser was set at 45$^{\circ}$ relative to the angles that excite the two fundamental modes of the fibre, both modes were excited equally. In both configurations, interference patterns were observed. The two spectra differ in that the positions of the maxima and minima are interchanged when comparing the crossed-polariser method to the aligned-polariser method. Example interference spectra  with the output polariser crossed plotted in green and aligned plotted in blue relative to the input polariser are shown in Fig.~\ref{Fig:MetBAnD}(b).

Group birefringence, B$_{g}$, depends on the fringe spacing in the measured spectral interference patterns. The fringe spacing, $\Delta\uplambda$, is the wavelength difference between adjacent peaks in the spectral interference pattern. The corresponding wavelength for each calculated value of B$_{g}$ can be taken as the centre wavelength between the adjacent peaks. The group birefringence as a function of wavelength can be calculated using:
\begin{align} \rm
B_{g}(\uplambda) = \frac{1}{L_{f}} \cdot \frac{\uplambda^{2}}{\Delta \uplambda},
\label{Eq:BgDelLam}
\end{align}
where L$_{f}$ is the length of the test fibre. We measured group birefringence of the four elliptical-core fibres. For each fibre, two sets of measurements were performed: one with the output polariser crossed with respect to the input polariser, and the other with the output polariser aligned relative to the input polariser.

\subsection*{Measurement of group velocity dispersion} 
Broadband dispersion measurements of the elliptical-core PM-ZBLAN fibres were performed using white-light spectral interferometry. The GVD measurement method is based on the approach presented in a previous studies~\cite{Mul02PbgDSpie,Hlu12DispJeos}. The experimental setup used for the GVD measurement is shown in Fig.~\ref{Fig:MetBAnD}(c) and is based on a Mach--Zehnder interferometer. After passing through the polariser (P$_{1}$, Thorlabs: LPMIR050-MP2), the light is split into two paths by a beam splitter (BS$_{1}$, Thorlabs: BSW511). The reflected part of the light is directed to the test arm of the interferometer, while the transmitted part enters the reference arm. In the test arm, a mirror M$_{1}$ directs the beam to an aspherical lens L$_{1}$, which couples the light into the fibre under test. The output from the fibre is collimated using an aspherical lens L$_{4}$, and the beam is then directed by mirrors M$_{2}$ and M$_{3}$, through the beam splitter (BS$_{2}$, Thorlabs: BSW511) and a second polariser (P$_{2}$, Thorlabs: LPMIR050-MP2). Finally, the light is coupled into the collection fibre (CF, Thorlabs: P1-23Z-FC-1) using an aspherical lens (L$_{5}$, Thorlabs: C036TME-D).

In the reference arm, light transmitted through the beam splitter BS$_{1}$ is focused using an aspherical lens L$_{2}$, which is identical to lens L$_{1}$. The beam is then collimated using lens L$_{3}$, which is identical to lens L$_{4}$. The collimated light is directed by mirrors M$_{4}$, M$_{5}$, and M$_{6}$ to a second beam splitter BS$_{2}$. The part of the light reflected from BS$_{2}$ passes through the polariser P$_{2}$ and is also coupled into the collection fibre using an aspherical lens L$_{5}$. Mirrors M$_{4}$ and M$_{5}$ are mounted on a one-dimensional translation stage, allowing the optical path length in the reference arm to be varied. This variable path length is indicated by the bidirectional arrow in Fig.~\ref{Fig:MetBAnD}(c). The setup is designed so that light in both the reference and test arms interacts with identical optical components, except for the fibre in the test arm. As a result, any difference in the properties of the light travelling through the two arms arises solely from the fibre under test. All mirrors used in the setup were silver coated mirrors (M$_{i}$, Thorlabs: PF10-03-P01), and all lenses were aspheric. The spectra from the collection fibre were measured using an optical spectrum analyser (Instrument Systems: Spectro 320).

To measure the dispersion of one of the fundamental modes of the fibre, polariser P$_{1}$ was rotated to selectively couple light into a single fundamental mode. Polariser P$_{2}$ was then aligned to transmit light corresponding to that same mode, thereby suppressing any residual light coupled into the orthogonal mode. When light from both arms is present in the measured spectrum, the optical path length in the reference arm can be adjusted using mirrors M$_{4}$ and M$_{5}$ mounted on a one-dimensional translation stage. This enables the optical path lengths in both the fibre arm and the reference arm to be made equal for a given wavelength. At this wavelength, the light accumulates the same phase in both arms; this wavelength region is referred to as the phase-matching wavelength. Spectral interference fringes appear on either side of this wavelength. An example of such a spectral interference pattern is shown in Fig.~\ref{Fig:MetBAnD}(d) in green, plotted with respect to the left-hand y-axis. The phase-matching wavelength occurs around 3.7~$\upmu$m. Due to a significant dip in the SC source around 1.7~$\upmu$m, the fringe visibility is low in that region, as it was not possible to couple equal amounts of light into both the fibre and reference arms.

To obtain the interference pattern shown in Fig.~\ref{Fig:MetBAnD}(d), the spectrum was first measured with light from both arms present simultaneously, denoted as I$_{Total}$. Spectra were then recorded from each arm individually while the other was blocked. The intensities from the individual arms were subtracted from I$_{Total}$, and the result was normalised to yield the spectral interference pattern I$_{Interf}$. The resulting interference fringes were then numbered, with the fringe number denoted by i, updated by 1 for each adjacent fringe maximum. When the numbering reaches the phase-matching wavelength, the direction of numbering was reversed. The fringe order of the example interference pattern, I$_{Interf}$, shown in Fig.~\ref{Fig:MetBAnD}(d), is plotted in blue with respect to the right-hand y-axis. The fringe order is then fitted to the equation:
\begin{equation} \label{Eq:Fring}
i = a_{1}\uplambda^{-5}+a_{2}\uplambda^{-3}+a_{3}\uplambda^{-1}+a_{4}\uplambda^{1}+a_{5}\uplambda^{3}-m,
\end{equation}
where, $a_{1}=-$A$_{1}$L$_{f}$, $a_{2}=-$A$_{2}$L$_{f}$, $a_{4}=-$A$_{4}$L$_{f}$, $a_{5}=-$A$_{5}$L$_{f}$, and m is an integer. The expression for the fringe order i, given in Eq.~(\ref{Eq:Fring}), is derived from the modified Cauchy equation for the refractive index~\cite{Hlu12DispJeos}. As the measurements are performed over a broad wavelength range within the transparent spectral region between the short- and long-wavelength material-resonance regions, this expression provides a suitable description of the wavelength dependence of the modal effective refractive index and hence of the measured fringe order. From the fit, the `a'-coefficients are obtained and subsequently converted to the `A'-coefficients. Using these A-coefficients, the measured GVD as a function of wavelength can be obtained. The group velocity dispersion D is given by:
\begin{equation} \label{Eq:Disp}
D = -\frac{1}{c}\big[20A_{1}\uplambda^{-5}+6A_{2}\uplambda^{-3}+2A_{4}\uplambda+12A_{5}\uplambda^{3}\big].
\end{equation}
A step-by-step derivation of the modified Cauchy expression, its relation to the spectral-interferometry fringe order, and the subsequent calculation of the dispersion parameter is available here~\cite{Shr20AndiThe} and further detailed in the supplementary material.

GVD measurements were repeated three times for each polarisation mode, with the phase-matching wavelength varied across the measurements. Note that the beam splitter shown in Fig.~\ref{Fig:MetBAnD}(a) was not required for the group birefringence measurements but was included from the outset to enable a seamless transition to the GVD measurements. The reflected light from the beam splitter was used for the group birefringence measurements. This approach avoided disturbing the fibre alignment when reconfiguring the setup from group birefringence to GVD measurements.
\section*{Data records}
The measured mean group birefringence and GVD data for all four elliptical-core PM-ZBLAN fibres are available through Figshare~\cite{EZbD25FigShr}. The wavelength versus mean group birefringence datasets for the four elliptical-core PM-ZBLAN fibres are provided as individual OpenDocument Spreadsheet (ODS) files. The wavelength versus GVD data are provided for each fibre, with separate ODS files for mode~I and mode~II.
\section*{Data overview}
\begin{figure}[htbp!]
\centering
\fbox{\includegraphics[width=0.98\linewidth]{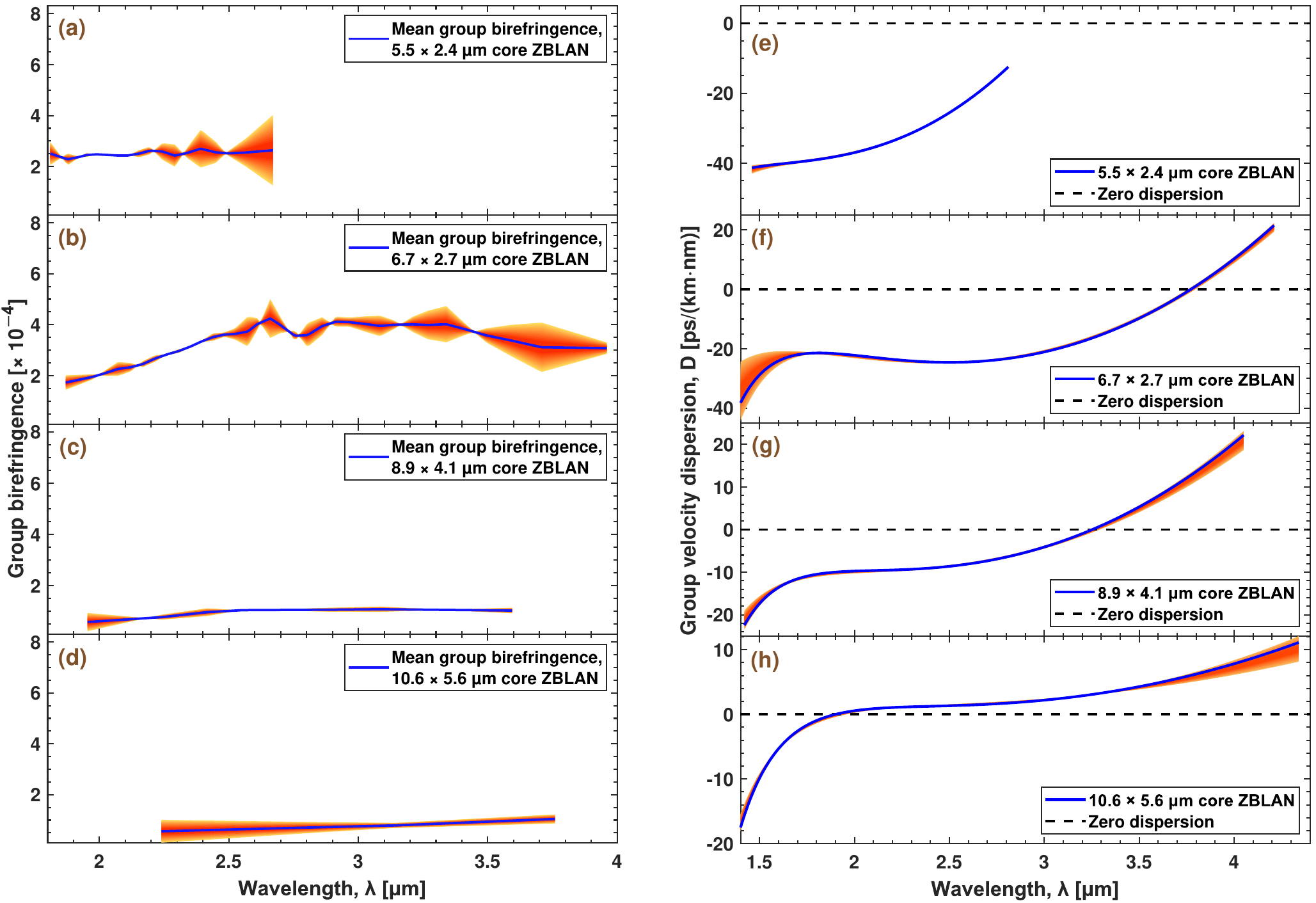}}
\caption{(a)--(d) Measured mean group birefringence values in blue for the four elliptical-core PM-ZBLAN fibres, along with the $\pm$2$\times$ combined standard uncertainty band shown in shades of orange. (e--h) Measured group velocity dispersion D curves for mode~I of the four elliptical-core PM-ZBLAN fibres, along with the $\pm$2$\times$ combined standard uncertainty error band shown in shades of orange. Blue curves correspond to one of the three independent measurements conducted for each fibre. The dashed black line indicates zero dispersion.}
\label{Fig:BirADisp}
\textcolor{color1}{\dotfill}
\end{figure}

To provide an overview of the measured datasets available through Figshare~\cite{EZbD25FigShr}, the mean group birefringence obtained from the crossed- and aligned-polariser measurements for each fibre is plotted in Figs.~\ref{Fig:BirADisp}(a)--(d).

The GVD of both fundamental modes of each fibre was measured. The mode with the lower absolute value of GVD at 2~$\upmu$m was designated as mode~I, and the orthogonal mode was designated as mode~II. The A-coefficients used to calculate the GVD for mode~I and mode~II of each fibre are provided in Table~\ref{Tab:Tab2}. These coefficients correspond to one of the three measurements performed for each mode rather than the mean of the three measurements. This is done because the GVD obtained from an individual measurement is a smooth curve, unlike the mean of the three measurements. The smooth curves can then be integrated or differentiated if needed. The GVD for this selected measurement of mode~I in the four fibres is plotted in Figs.~\ref{Fig:BirADisp}(e)--(h) in blue. Uncertainties in the measured GVD will propagate when $\upbeta_{2}(\upomega)$ is integrated to obtain $\upbeta_{1}(\upomega)$ and subsequently $\upbeta(\upomega)$. When $\upbeta_{1}(\upomega)$ is obtained by numerically integrating the measured $\upbeta_{2}(\upomega)$, and $\upbeta(\upomega)$ is subsequently obtained by numerically integrating $\upbeta_{1}(\upomega)$, the propagation of uncertainty is linear. Since the fit to the fringe order was performed with wavelength in $\upmu$m, the variable $\uplambda$ in Eq.~(\ref{Eq:Disp}) must also be expressed in $\upmu$m, while the speed of light is expressed in m/s. Table~\ref{Tab:Tab2} also lists the measurement ranges, which are the wavelength range over which the interference fringes were measured. Using Eq.~(\ref{Eq:Disp}), the GVD can also be estimated outside, but close to, this range with reasonable accuracy.
\begin{table}[htbp!]
\centering
\setlength{\extrarowheight}{0pt}
\setlength{\tabcolsep}{8pt} 
\begin{NiceTabular}{c l S S S S c}[cell-space-limits=0pt]
\CodeBefore
\rowcolor{color3}{1}
\Body
\toprule
\textbf{Fibre [$\upmu$m]} & \textbf{Mode} & \textbf{A\textsubscript{1}} & \textbf{A\textsubscript{2}} & \textbf{A\textsubscript{4}} & \textbf{A\textsubscript{5}} & \textbf{Measurement range} \\
\midrule
\Block{2-1}{5.5 $\times$ 2.4} & Mode I & 504.245 & 131.132 & 4740.566 & -86.368 & \Block{2-1}{1.460--2.810 $\upmu$m} \\
\cmidrule(lr){2-6}
& Mode II & -1037.736 & 3270.755 & 4125.000 & -74.764 & \\
\midrule
\Block{2-1}{6.7 $\times$ 2.7} & Mode I & 6717.172 & -9969.697 & 3406.566 & -38.449 & \Block{2-1}{1.400--4.210 $\upmu$m} \\
\cmidrule(lr){2-6}
& Mode II & 3964.646 & -6121.212 & 3251.010 & -37.763 & \\
\midrule
\Block{2-1}{8.9 $\times$ 4.1} & Mode I & 4013.333 & -5323.810 & 1703.810 & -25.133 & \Block{2-1}{1.420--4.050 $\upmu$m} \\
\cmidrule(lr){2-6}
& Mode II & 4581.905 & -5990.476 & 1798.095 & -26.562 & \\
\midrule
\Block{2-1}{10.6 $\times$ 5.6} & Mode I & 3739.196 & -4276.382 & 318.492 & -5.945 & \Block{2-1}{1.400--4.340 $\upmu$m} \\
\cmidrule(lr){2-6}
& Mode II & 4978.392 & -5954.774 & 484.573 & -7.884 & \\
\bottomrule
\end{NiceTabular}
\caption{A-coefficients to obtain the measured group velocity dispersion D for mode~I and mode~II of each elliptical-core PM-ZBLAN fibre. The measurement range indicates the wavelength interval over which spectral interference fringes were measured.}
\label{Tab:Tab2}
\textcolor{color1}{\dotfill}
\end{table}

For the 5.5~$\times$~2.4~$\upmu$m core fibre, interference fringes could only be recorded up to 2.81~$\upmu$m in our setup. This limitation affected both the GVD and group birefringence measurements for this fibre and is primarily attributed to the small core size and relatively low NA, which increase confinement loss at longer wavelengths and prevent efficient guidance of longer wavelengths in the free-space coupling configuration used for the broadband light source.

The GVD of the 5.5~$\times$~2.4~$\upmu$m core fibre is normal across the entire wavelength range of the measurement. The 6.7~$\times$~2.7~$\upmu$m core fibre has a ZDW at 3.772~$\upmu$m, the 8.9~$\times$~4.1~$\upmu$m core fibre has a ZDW at 3.252~$\upmu$m, and the 10.6~$\times$~5.6~$\upmu$m core fibre has a ZDW at 1.9~$\upmu$m. These ZDWs correspond to mode~I of the measurements plotted in blue in Figs.~\ref{Fig:BirADisp}(e)--(h).
\section*{Technical validation}
The experimental methods for measuring group birefringence and GVD are based on earlier studies and represent well-established techniques that have been widely employed for group birefringence~\cite{Fol04LmaOe} and GVD~\cite{Ci18DispJosaB,Rao22PmNpm} measurements in the NIR, including for PM fibres. The measured normal-dispersion profiles of the 6.7~$\times$~2.7~$\upmu$m and 8.9~$\times$~4.1~$\upmu$m core PM-ZBLAN fibres have also supported the demonstration of ultra-low-noise SC generation in normal-dispersion ZBLAN fibres pumped at 1.85~$\upmu$m~\cite{ShrUlZ26Ol}. Similar experimental setups have been used to measure GVD in the 1.3--4.5~$\upmu$m range in chalcogenide fibres~\cite{Chr14ChaNp} and in waveguides on gallium lanthanum sulphide~\cite{Dem16GioOe}. To quantify the experimental uncertainty, an error analysis was carried out for both the group birefringence and GVD measurements.
\subsection*{Calculation of error}
The total uncertainty in the group birefringence measurements was calculated by combining two independent sources of uncertainty using the root-sum-of-squares method: experimental variability and the uncertainty in the measured fibre length. For each fibre, two separate measurements were performed using complementary techniques--one with the transmission axis of the output polariser crossed relative to the input polariser, and the other with the two polarisers aligned. The two resulting datasets were interpolated to a common wavelength grid, which enabled a direct point-by-point calculation of the mean group birefringence and the standard deviation between the two measurements.
\begin{table}[htbp!]
\centering
\setlength{\extrarowheight}{0pt}
\setlength{\tabcolsep}{20pt}
\begin{NiceTabular}{c >{\centering\arraybackslash}m{2.7cm} >{\centering\arraybackslash}m{2.7cm} >{\centering\arraybackslash}m{2.7cm}}[cell-space-limits=0pt]
\CodeBefore
\rowcolor{color3}{1,2}
\Body
\toprule
\Block{2-1}{\textbf{\makecell{Fibre core\\{[$\upmu$m]}}}} & \multicolumn{3}{c}{\textbf{[Mean group birefringence $\pm$ 2 $\times$ combined standard uncertainty] $\times$ 10$^{-4}$}} \\
\cmidrule(lr){2-4}
& \textbf{at 2.00 $\upmu$m} & \textbf{at 2.40 $\upmu$m} & \textbf{at 2.80 $\upmu$m} \\
\midrule
5.5 $\times$ 2.4 & 2.47 $\pm$ 0.03 & 2.68 $\pm$ 0.69 & --- \\
\midrule
6.7 $\times$ 2.7 & 2.03 $\pm$ 0.02 & 3.37 $\pm$ 0.04 & 3.59 $\pm$ 0.42 \\
\midrule
8.9 $\times$ 4.1 & 0.62 $\pm$ 0.28 & 0.95 $\pm$ 0.15 & 1.06 $\pm$ 0.05 \\
\midrule
10.6 $\times$ 5.6 & --- & 0.60 $\pm$ 0.38 & 0.71 $\pm$ 0.22 \\
\bottomrule
\end{NiceTabular}
\caption{Mean group birefringence values for the four elliptical-core PM-ZBLAN fibres at selected wavelengths, with their corresponding $\pm$2 $\times$ combined standard uncertainty.}
\label{Tab:Tab3}
\textcolor{color1}{\dotfill}
\end{table}
In addition to this experimental variability, the uncertainty from the measured fibre length was also included. An uncertainty of $\pm$1~mm in the fibre length was taken into account to calculate the corresponding proportional standard uncertainty contribution to the group birefringence. This length-dependent contribution was then combined with the standard deviation from the two measurements using a root-sum-of-squares method to obtain the combined standard uncertainty.

The mean group birefringence for the four elliptical-core PM-ZBLAN fibres is shown in blue in Fig.~\ref{Fig:BirADisp}(a)--(d), with the associated uncertainty band shown in shades of orange. The uncertainty band represents $\pm$2 times the combined standard uncertainty. The group birefringence values obtained from the crossed-polariser and aligned-polariser methods intersect at certain wavelengths, which leads to reduced uncertainty at those wavelengths. The mean group birefringence at selected wavelengths for the fibres is provided in Table~\ref{Tab:Tab3}, with a corresponding uncertainty of $\pm$2$\times$ combined standard uncertainty.
\begin{table}[htbp!]
\centering
\setlength{\extrarowheight}{0pt}
\setlength{\tabcolsep}{12pt}
\begin{NiceTabular}{c c c c c c}[cell-space-limits=0pt]
\CodeBefore
\rowcolor{color3}{1,2}
\Body
\toprule
\Block{2-1}{\textbf{\makecell{Fibre core\\{[$\upmu$m]}}}}
& \Block{2-1}{\textbf{Mode}} & \multicolumn{4}{c}{\textbf{[Mean GVD $\pm$ 2 $\times$ combined standard uncertainty]  [ps$\mathbin{/}$(km$\cdot$nm)]}}\\
\cmidrule(lr){3-6}
& & \textbf{at 1.55 $\upmu$m} & \textbf{at 2.00 $\upmu$m} & \textbf{at 2.80 $\upmu$m} & \textbf{at 4.00 $\upmu$m} \\
\midrule
\Block{2-1}{5.5 $\times$ 2.4} & Mode I & $-$40.81 $\pm$ 0.68 & $-$36.93 $\pm$ 0.36 & $-$12.96 $\pm$ 0.27 & ---\\
\cmidrule(lr){2-6}
& Mode II & $-$40.05 $\pm$ 4.54 & $-$37.43 $\pm$ 0.66 & $-$13.73 $\pm$ 3.20 & --- \\
\midrule
\Block{2-1}{6.7 $\times$ 2.7} & Mode I & $-$24.71 $\pm$ 3.61 & $-$22.49 $\pm$ 0.67 & $-$23.21 $\pm$ 0.36 & $+$09.90 $\pm$ 1.08 \\
\cmidrule(lr){2-6}
& Mode II & $-$23.79 $\pm$ 1.58 & $-$24.48 $\pm$ 0.51 & $-$23.45 $\pm$ 0.29 & $+$11.33 $\pm$ 0.82 \\
\midrule
\Block{2-1}{8.9 $\times$ 4.1} & Mode I & $-$14.89 $\pm$ 0.73 & $-$09.90 $\pm$ 0.36 & $-$06.41 $\pm$0.32 & $+$19.15 $\pm$ 2.07 \\
\cmidrule(lr){2-6}
& Mode II & $-$14.79 $\pm$ 5.02 & $-$10.14 $\pm$ 0.37 & $-$06.45 $\pm$ 0.35 & $+$20.50 $\pm$ 2.77 \\
\midrule
\Block{2-1}{10.6 $\times$ 5.6} & Mode I & $-$07.19 $\pm$ 0.25 & $+$00.45 $\pm$ 0.21 & $+$01.82 $\pm$ 0.20 & $+$07.25 $\pm$ 1.14 \\
\cmidrule(lr){2-6}
& Mode II & $-$08.68 $\pm$ 1.43 & $+$00.25 $\pm$ 0.63 & $+$01.50 $\pm$ 0.25 & $+$08.37 $\pm$ 0.79 \\
\bottomrule
\end{NiceTabular}
\caption{Mean GVD values for the two fundamental modes of each of the four elliptical-core PM-ZBLAN fibres at selected wavelengths, with their corresponding $\pm$2$\times$ combined standard uncertainty.}
\label{Tab:Tab4}
\textcolor{color1}{\dotfill}
\end{table}

For the GVD measurements, three acquisitions were performed for each fibre and mode by varying the phase-matching wavelength while keeping the fibre alignment unchanged. These acquisitions were used to quantify the sensitivity of the obtained GVD to the choice of phase-matching wavelength, rather than as fully independent repetitions of the complete experimental procedure. Each acquisition was analysed independently, starting from the measured spectral interference pattern and ending with a GVD curve. Changing the phase-matching wavelength changes the measured fringe-order curve, which leads to variations in the fitted curve derived from the modified Cauchy equation and, consequently, in the GVD obtained from the measurement.

At each wavelength, the mean and standard deviation of the three retrieved GVD curves were calculated. This standard deviation was treated as the standard uncertainty contribution associated with varying the phase-matching wavelength. It includes the combined effects of changes in the fringe-order curve, uncertainty in identifying fringe peaks and valleys, and short-term OSA measurement noise. The uncertainty arising from the measured fibre length was included separately. An uncertainty of $\pm$1~mm in the fibre length was used to calculate the corresponding proportional standard uncertainty contribution to the GVD. This contribution was combined with the standard uncertainty from varying the phase-matching wavelength using a root-sum-of-squares method to obtain the combined standard uncertainty.

A representative GVD curve is shown in blue in Fig.~\ref{Fig:BirADisp}(e)--(h), with the corresponding uncertainty band shown in shades of orange. The uncertainty band represents $\pm2$ times the combined standard uncertainty. The mean GVD values at selected wavelengths for both fundamental modes of each fibre are provided in Table~\ref{Tab:Tab4}, together with the corresponding $\pm2$ combined standard uncertainty.
\section*{Data availability}
The wavelength versus measured mean group birefringence data for the four elliptical-core PM-ZBLAN fibres are available through Figshare~\cite{EZbD25FigShr}. The measured GVD data for both fundamental modes of each fibre are also available through Figshare~\cite{EZbD25FigShr}. The A-coefficients used in Eq.~(\ref{Eq:Disp}) to calculate the GVD curves are provided in Table~\ref{Tab:Tab2}.
\section*{Code availability}
Standard software packages were used for all analyses. No additional code is available.
\bibliography{SciDatZb25}
\section*{Acknowledgements}
The authors thank FiberLabs Inc., Japan, for providing the ZBLAN fibre samples, the data presented in Table~\ref{Tab:Tab1}, and the wavelength-dependent core and cladding refractive indices shown in Fig.~\ref{Fig:SimDisp}(a).
\section*{Author contributions}
S.R.D.S. and C.R.P. designed, and set up the group birefringence and dispersion measurement system. S.R.D.S. performed the experiments. S.R.D.S., A.R., and C.R.P. carried out the data processing and analysis. O.B. and A.M. oversaw the experimental work. All authors participated in the planning, drafting, and revision of the manuscript.
\section*{Funding}
Danmarks Frie Forskningsfond project no. 2031-00009B; VILLUM Fonden (2021 Villum Investigator project no. 00037822: Table-Top Synchrotrons); Innovation Fund Denmark project no. 2105-00039B (HYPERSORT); EU's Horizon Europe project no. 101058054 (TURBO); Schweizerischer Nationalfonds zur F\"{o}rderung der Wissenschaftlichen Forschung (TMPFP2\_210543, PCEFP2\_181222).
\section*{Competing interests}
The authors declare no competing interests.
\section*{Additional information}
\textbf{Supplementary information} Supplementary information accompanies this manuscript.\\

\noindent
\textbf{Correspondence} and requests for materials should be addressed to S.R.D.S. \\

\noindent
\textcopyright~The Author(s) 2025
\end{document}